\documentclass[%
 aip,
 amsmath,amssymb,
preprint,%
]{revtex4-1}

\usepackage{graphicx}
\usepackage{dcolumn}
\usepackage{bm}
\usepackage[utf8]{inputenc}
\usepackage[T1]{fontenc}
\usepackage{mathptmx}
\usepackage{epstopdf}
\usepackage{multirow}
\usepackage[dvipsnames]{xcolor}

\begin{document}

\preprint{AIP/123-QED}

\title[Fan-out enabled spin wave majority gate]{Fan-out enabled spin wave majority gate}
\author{Abdulqader Mahmoud}
\email{A.N.N.Mahmoud@tudelft.nl}
\affiliation{ 
Delft University of Technology, Mekelweg 2, 2628 CD Delft, The Netherlands 
}%
\author{Frederic Vanderveken}%
\author{Christoph Adelmann}
\author{Florin Ciubotaru}
\affiliation{ 
IMEC, Kapeldreef 75, B-3001 Leuven, Belgium
}%

\author{Said Hamdioui$^1$}
\author{Sorin Cotofana$^1$}

\begin{abstract}
By its very nature, Spin Wave (SW) interference provides intrinsic support for Majority logic function evaluation. Due to this and the fact that the $3$-input Majority (MAJ3) gate and the Inverter constitute a universal Boolean logic gate set, different MAJ3 gate implementations have been proposed. However, they cannot be directly utilized for the construction of larger SW logic circuits as they lack a key cascading mechanism, i.e., fan-out capability. In this paper, we introduce a novel ladder-shaped SW MAJ3 gate design able to provide a maximum fan-out of $2$ (FO2). The proper gate functionality  is validated by means of micromagnetic simulations, which also demonstrate that the amplitude mismatch between the two outputs is negligible proving  that an FO2 is properly achieved. Additionally, we evaluate the gate area and compare it with SW state-of-the-art and $15$nm CMOS counterparts working under the same conditions. Our results indicate that the proposed structure requires $12$x less area than the $15$ nm CMOS MAJ3 gate and that at the gate level the fan-out capability results in $16$ \% area savings, when compared with the state-of-the-art SW majority gate counterparts.

\end{abstract}

\maketitle

The rapid increase of available row data led to an abrupt downscaling of the CMOS technology in order to meet the continuously increasing application demand for high performance computation platforms \cite{data1}. However, CMOS scaling became more and more difficult due to various technological hurdles such as: (i) quantum mechanics related phenomena and physical limitations such as leakage \cite{cmosscaling2}, (ii) high failure rate and short life time of devices \cite{cmosscaling1}, and (iii) steep fabrication cost increase not justifiable by scaling economical benefits \cite{cmosscaling2}. As a result, different emerging technologies are now explored as potential candidates for future partial/total CMOS replacement \cite{survey1,survey2}. One of them relies on Spin Waves (SW) interference within magnetic waveguides \cite{survey1,survey2}. Preliminary investigations suggest that SW based computing potentially enables ultra low power consumption at acceptable delay and has great scalability potential \cite{survey1,survey2}. SW computing is based on wave interference, which can be either constructive or destructive depending on the interfering SWs phases. This principle is used to build SW logic gates. Spin wave interferometer, e.g., Mach-Zhender interferometer was used to investigate this phenomena \cite{ref5,ref6,ref7,ref8,ref9}. To this end, different logic and Majority gate designs were introduced \cite{logic1,Excitation_table_ref16,logic13,logic14,logic20,logic9,logic101,ref1,ref2,ref3,ref4} they all, with the exception of \cite{logic9,logic101}, make use of bent waveguides through which weak signals as SWs do not properly propagate and attenuate very fast. 

As $3$-input Majority gate (MAJ3) together with Inverter form a universal Boolean logic gate set, they provide the foundation for the potential implementation of complex SW circuits \cite{logic1}. However, building larger circuits requires gates with fanout capability, which none of the previously mentioned designs posses. Thus, if a certain Majority gate has to provide its output to more than one gate input, it has to be replicated. For example, if a gate output has a fanout $f > 1$, all the gates on its cone of influence starting for the circuit primary inputs have to be replicated $f$ times. Given that practical circuits include many such gates the lack of fanout capability results in substantial area and energy consumption overheads.  The SW circuit fanout issue has been addressed and by magnonic splitters \cite{fanout1,fanout2,logic24,ref10} or caustic beams \cite{fanout1} based solutions have been proposed. However, the presented designs require large frequency bands and are not scalable. If the magnetic field is applied in plane, the T-shape magnonic splitter \cite{fanout2} relies on SW mode (backward volume and surface) conversion. Given that the dispersion relation is magnetic field direction dependent, such an approach results in complex SW interference patterns, which precludes the utilization of T-shape magnonic splitters in the design of large SW circuits. The possibility to implement a magnonic splitter by voltage controlled reconfigurable nano-channels was discussed in \cite{logic24}, however, no detailed analysis of the spin wave quality after splitting has been provided. Additionally, a nonlinear directional coupler that allows SW transmission from a waveguide to another was investigated \cite{ref10} and demonstrated the SW power dependency of this phenomenon. However, this concept splits the SW energy and cannot provide SW replication, which is crucial for gate fanout achievement.

In view of the above, it can be concluded that SW based computing with potential ultra low energy consumption cannot become reality without gate intrinsic fanout capabilities. Here, we overcome this challenge and introduce a generic SW Majority gate structure that provides natural fanout support. Our structure is based on an area efficient $3$-input Majority ladder-shaped SW gate structure that is able to provide a maximum fanout of $2$. This concept has been validated by means of micromagnetic simulations with the Object Oriented Micromagnetic Framework (OOMMF). 

Generally speaking, the proposed gate can operate with any SW type, however, each SW type has its proper dispersion relation, which plays a crucial role in the actual gate design. Magnetostatic Spin Waves (MSW) can be classified into three limiting cases: Magnetostatic Surface Spin Wave (MSSW), Backward Volume Magnetostatic Spin Wave (BVMSW), and Forward Volume Magnetostatic Spin Wave (FVMSW) \cite{Magnetostatics_ref3}. Depending on the wave propagation direction, BVMSW and MSSW exhibit different dispersion relations. This complicates the circuit design because similar SW propagation in both horizontal and vertical directions is required. For FVMSWs, which propagate in a perpendicular plane to the static magnetization orientation, SW exhibit the same dispersion relation regardless of the wave vector orientation. In this view, we rely on them in the gate design introduced in the following lines.

Different SW excitation (and detection) methods exist, e.g., microstrip antennas \cite{ref101,Magnonic_crystals_for_data_processing}, magnetoelectric cells \cite{ excitation1,excitation2,excitation2,excitation3}, spin orbit torque \cite{ref100,excitation4}. A spin wave propagates through the waveguide with a wavelength $\lambda$, frequency $f$, amplitude $A$, and phase $\phi$. Information can be encoded in its amplitude, phase, or both of them. If multiple SWs coexist in a waveguide, the computation can be performed using wave interference. Two waves with the same $\lambda$, $A$, and $f$ can interfere constructively or destructively depending on their relative phase difference: (i) in-phase SWs interfere constructively and the resulting wave has doubled amplitude, (ii) out-of-phase SWs interfere destructively, and therefore cancel each other. If more than $2$ equal $\lambda$ and $f$ SWs interfere, the result reflects a Majority decision, i.e., if more SW have  $\phi = \pi$ (logic "1") than $\phi = 0$ (logic "0"), the resultant SW has $\phi=\pi$, and $\phi=0$ otherwise. This means that SW interference provides natural support for direct (no Boolean gates are required) Majority gate implementations. For example, a CMOS implementation of a $3$-input Majority gate requires $18$ transistors whereas a single magnetic waveguide is enough for the SW counterpart \cite{logic1,logic9}. In the linear regime, it is possible to have simultaneous propagation of spin waves with different frequencies. The information can be encoded in the phase of the spin wave at each and every frequency, therefore, SW gates inherently enable parallel computation on shared hardware resources. Additionally, if the involved waves have different amplitude, they still constructively or destructively interfere depending on phase difference. However, this generates multiple SWs with different amplitude values, which could be beneficial for the realization of multi-valued logic gates. In the most general case, SWs with different amplitudes, phases, wavelengths, and frequencies can be excited and intricately interfere in the same waveguide. This provides promising alternative avenues towards novel, yet to be discovered, SW based computing paradigms and systems.

\begin{figure}[t]
\centering
  \includegraphics[width=0.3\linewidth]{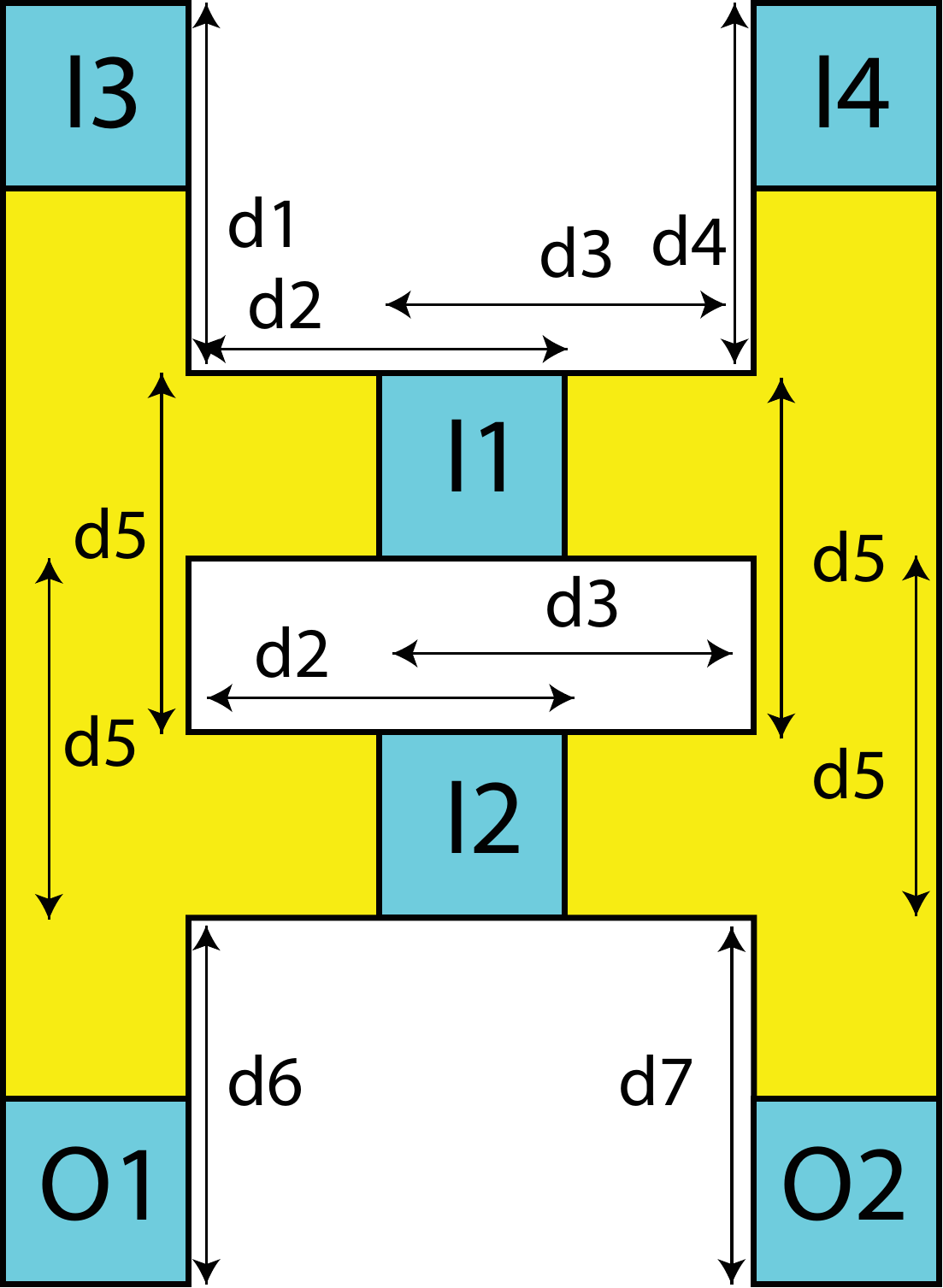}
  \caption{$3$-input Majority Gate with Fanout Capability.}
  \label{fig:system}
\end{figure}

In this paper, we propose  a $3$-input Majority gate (MAJ3) that has a ladder-shape structure, as depicted in Figure \ref{fig:system}. The inputs are excited at ($I_1$, $I_2$, $I_3$, $I_4$) and the outputs are read from ($O_1$, $O_2$). 

To obtain a proper interference pattern at the crosspoints, the waveguide width $w$ has to be less than or equal to the wavelength $\lambda$. Also, the excited SWs should have the same amplitude $A$. In addition, all excited SWs are required to have the same frequency to achieve the desired interference pattern. We propose a generic device layout, its dimensions and some critical distances $d_{i}$ (where i=1,2,\ldots,7) are expressed in terms of spin wave wavelengths as indicated in Figure \ref{fig:system}. For example, if $\lambda$ wavelength SWs have to constructively interfere when they have the same phase and destructively otherwise, $d_1$, $d_2$, $d_3$, $d_4$, and $d_5$ must be equal with $n\lambda (n=1,2,3, \ldots)$. If the opposite behaviour is targeted, $d_1, d_2, d_3, d_4$ and $d_5$ must be equal with $\frac{n}{2} \lambda (n=1,3,5, \ldots)$. Moreover, to obtain a proper fanout of $2$, i.e., outputs with the same energy levels, the structure has to be symmetric, thus $d_1$ to $d_5$ must have the same value. 

In contrast with CMOS gates, SW gates can provide both direct and inverted output by properly adjusting the output transducer position versus the output interference point. In this way the direct and inverted result can be read at a distance of $n \lambda (n=1,2,3, \ldots)$ and  $\frac{n}{2} \lambda (n=1,2,3, \ldots)$ from the last interference, respectively. In our case, MAJ(a,b,c) and $\overline{MAJ}(a,b,c)$ are obtained at $d_6$ = $d_7$ =  $n\lambda (n=1,2,3,\ldots)$ and $d_6$ = $d_7$  =  $(\frac{n}{2} \lambda (n=1,3,5, \ldots)$, respectively, and both outputs exhibit the same energy because of the structure symmetry. 

Intuitively speaking, the Majority gate operates as follows: (i) SWs with appropriate phases are initiated at $I_1$, $I_2$, $I_3$, and $I_4$ to the targeted logic value ($0$ or $1$). (ii) The excited SWs propagate (in both directions in the horizontal and vertical waveguides) and interfere when meeting each other. The resulting wave propagates toward the outputs $O_1$ and $O_2$. Thanks to the symmetry of the device and the isotropic behavior of the spin waves in this configuration, the waves arriving at the gate outputs are identical, thus, the $3$-input Majority gate exhibits a fanout of $2$. It is worth-mentioning that $I_3$ has effect on $O_2$ as spin-wave signal excited at $I_3$ propagates through $I_1$ and $I_2$. Also, $I_4$ has effect on $O_1$ as spin-wave signal excited at $I_4$ propagates through $I_1$ and $I_2$. In addition, spin wave excited at $I_1$ and $I_2$ face edges while its propagation to the output, in contrast to $I_3$ and $I_4$, which have straight path to the outputs. Therefore, $I_3$  and $I_4$ are excited at lower energy than $I_1$ and $I_2$ as will be discussed further later in this paper.

It is worth-mentioning that $I_3$ has effect on $O_2$ as the SW excited at $I_3$ propagates through $I_1$ and $I_1$. Similarly, $I_4$ has effect on $O_1$ spin-wave signal excited at $I_4$ propagates through $I_1$ and $I_2$. In addition, SWs excited at $I_1$ and $I_2$ face edges while they propagate towards the outputs while $I_3$ and $I_4$ generated SWs have straight path to $O_1$ and $O_2$, respectively. Therefore, $I_3$ and $I_4$ are excited at lower energy than $I_1$ and $I_2$ as further discussed in the paper.

\begin{table}
\caption{MAJ3 Truth Table.}
\label{table:1}
  \begin{tabular}{|c|c|c|c|c|c|c|c|c|}
    \hline
    $I_1$ & $I_2$ & $I_3$ & $O_1$ & $I_1$ & $I_2$ & $I_4$ & $O_2$ & Indication in Figure 2\\
    \hline
    $0$ & $0$ & $0$ & $0$ & $0$ & $0$ & $0$ & $0$ & (i) \\
    \hline
    $0$ & $0$ & $1$ & $0$ & $0$ & $0$ & $1$ & $0$ & (ii) \\
    \hline
    $0$ & $1$ & $0$ & $0$ & $0$ & $1$ & $0$ & $0$ & (iii)\\
    \hline
    $0$ & $1$ & $1$ & $1$ & $0$ & $1$ & $1$ & $1$ & (iv)\\
    \hline
    $1$ & $0$ & $0$ & $0$ & $1$ & $0$ & $0$ & $0$ & (v)\\
    \hline
    $1$ & $0$ & $1$ & $0$ & $1$ & $0$ & $1$ & $0$ & (vi)\\
    \hline
    $1$ & $1$ & $1$ & $1$ & $1$ & $1$ & $0$ & $1$ & (vii)\\
    \hline
    $1$ & $1$ & $1$ & $1$ & $1$ & $1$ & $1$ & $1$ & (viii)\\
    \hline
  \end{tabular}
\end{table} 

\begin{figure}
\centering
  \includegraphics[width=0.8\linewidth]{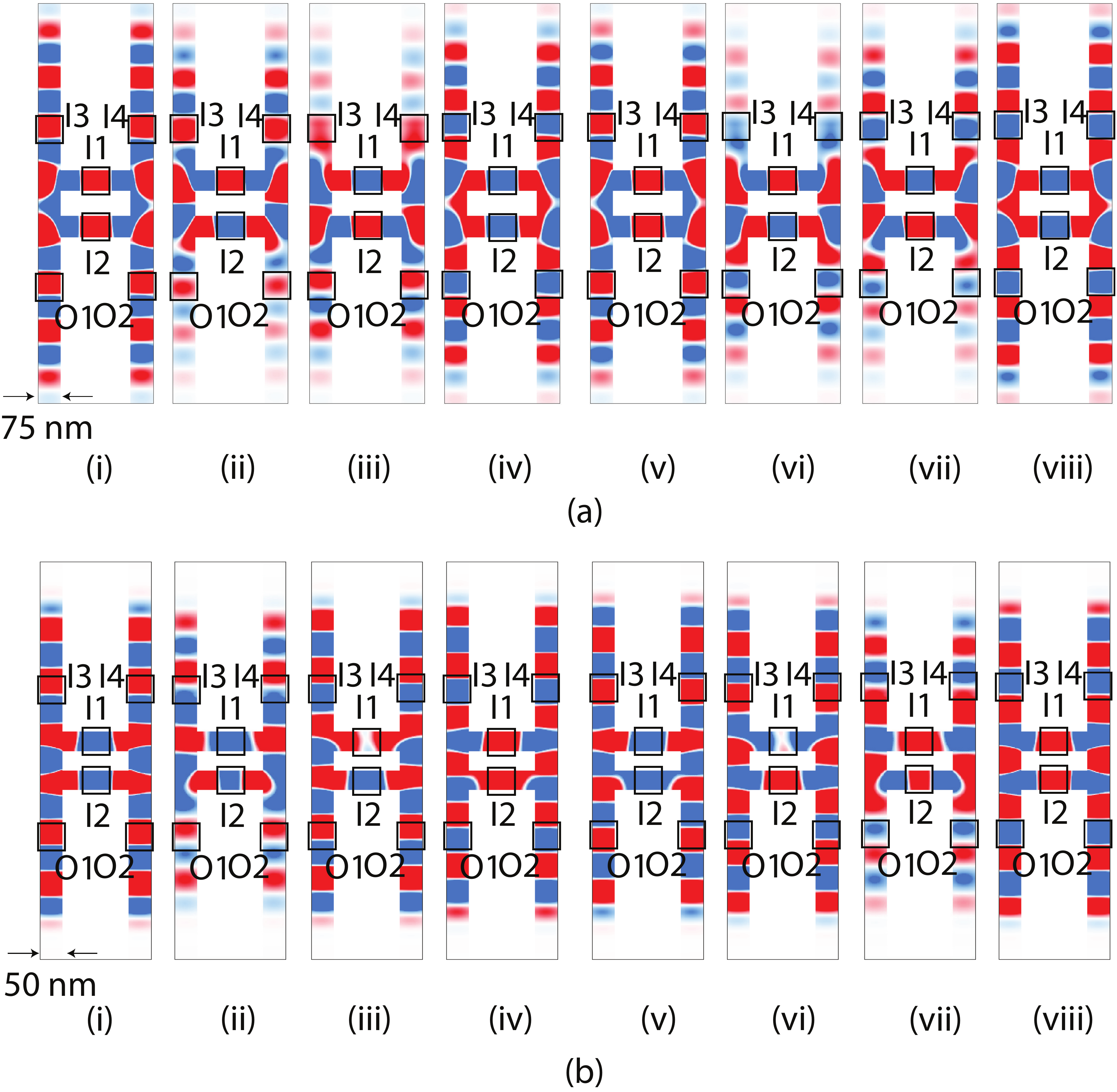}
  \caption{Color coded snapshots of the magnetization state demonstrating all majority functions for two width of SW waveguide: (a) $75$ nm and (b) $50$ nm. Blue represents logic $1$, which presents a phase of $\pi$, red presents logic $0$ which presents phase $0$, the input order is ($I_3$ $I_2$ $I_1$) and ($I_4$ $I_2$ $I_1$), and  i), ii), iii), iv), v), vi), vii), and viii) present the gate reaction to  (0 0 0),  (0 0 1), (0 1 0), (0 1 1), (1 0 0), (1 0 1), (1 1 0), and (1 1 1) input patterns, respectively.}
  \label{fig:results}
\end{figure} 

We validate the proposed majority gate by means of micromagnetic simulations while making use of $Fe_{60}Co_{20}B_{20}$ waveguides, with a Perpendicular Magnetic Anisotropy (PMA) field greater than the magnetic saturation, which means that no external magnetic field is required for  proper gate operation. We instantiated a MAJ3 gate for waveguide width $w = 75$ nm, and to simplify the interference pattern, we selected a larger wavelength than $w$, SW wavelength $\lambda = 165$ nm, which implies that  $d_1 = d_2 = d_3 = d_4 = d_5 = d_6 = d_7= 165$ nm. Further, we assume the following values of the relevant parameters~\cite{parameters}: magnetic saturation $M_s = 1.1 \times 10^6$  $A/m$, exchange stiffness $A_{exch} = 18.5$ $pJ/m$, damping constant $\alpha = 0.004$, perpendicular anisotropy constant $k_{ani} = 8.3177 \times 10^5$ $J/m^3$, and waveguide thickness $t = 1$ nm.  We calculated the FVMSW dispersion relation for these parameters, and for $\lambda =165$ nm, and $k=2\pi/\lambda=38$ rad/$\mu$m,  the SW frequency is determined to be $f = 6.5$ GHz. To get some indication of the MAJ3 scaling implications, we also designed smaller structures, e.g., $w=50$ nm, with $\lambda=110$ nm and $f=9$ GHz. This makes the distances $d_1 = d_2 = d_3 = d_4 = d_5 = d_6 = d_7= 110$ nm.

The proposed design combines two Majority gates operating in parallel on the same input set as it can be observed in Table \ref{table:1}. $I_1$, $I_2$, and $I_3$ constitute the first Majority gate with its output being detected at $O_1$, whereas $I_1$, $I_2$, and $I_4$ constitute the second Majority gate with $O_2$ as output. Figure \ref{fig:results} presents OOMMF simulation results for the proposed $w=75$ nm and $w=50$ nm MAJ3 gates, under all possible input combinations. Note that in the Figure  blue presents logic $"1"$ (i.e., phase of $\pi$), red presents logic $"0"$ (i.e., phase $0$), the input order is ($I_3$ $I_2$ $I_1$) and ($I_4$ $I_2$ $I_1$), and  i), ii), iii), iv), v), vi), vii), and viii) captures the gate reaction to  (0 0 0),  (0 0 1), (0 1 0), (0 1 1), (1 0 0), (1 0 1), (1 1 0), and (1 1 1) input patterns, respectively. As it can be observed from Figure \ref{fig:results}, the results are in agreement with the MAJ3 true table in Table \ref{table:1}. If $I_1$ = $I_2$ = $I_3$ = $0$ or the majority of the inputs are $0$ then $O_1$ = $O_2$ = $0$ (red), whereas if the majority of the inputs are $1$, then the outputs $O_1$ and $O_2$ are $1$ (blue), as expected. In addition, it can be noticed in Figure \ref{fig:results} that the scaling doesn't affect the functionality of the Majority gate. 

\begin{figure}[t]
\centering
  \includegraphics[width=\linewidth]{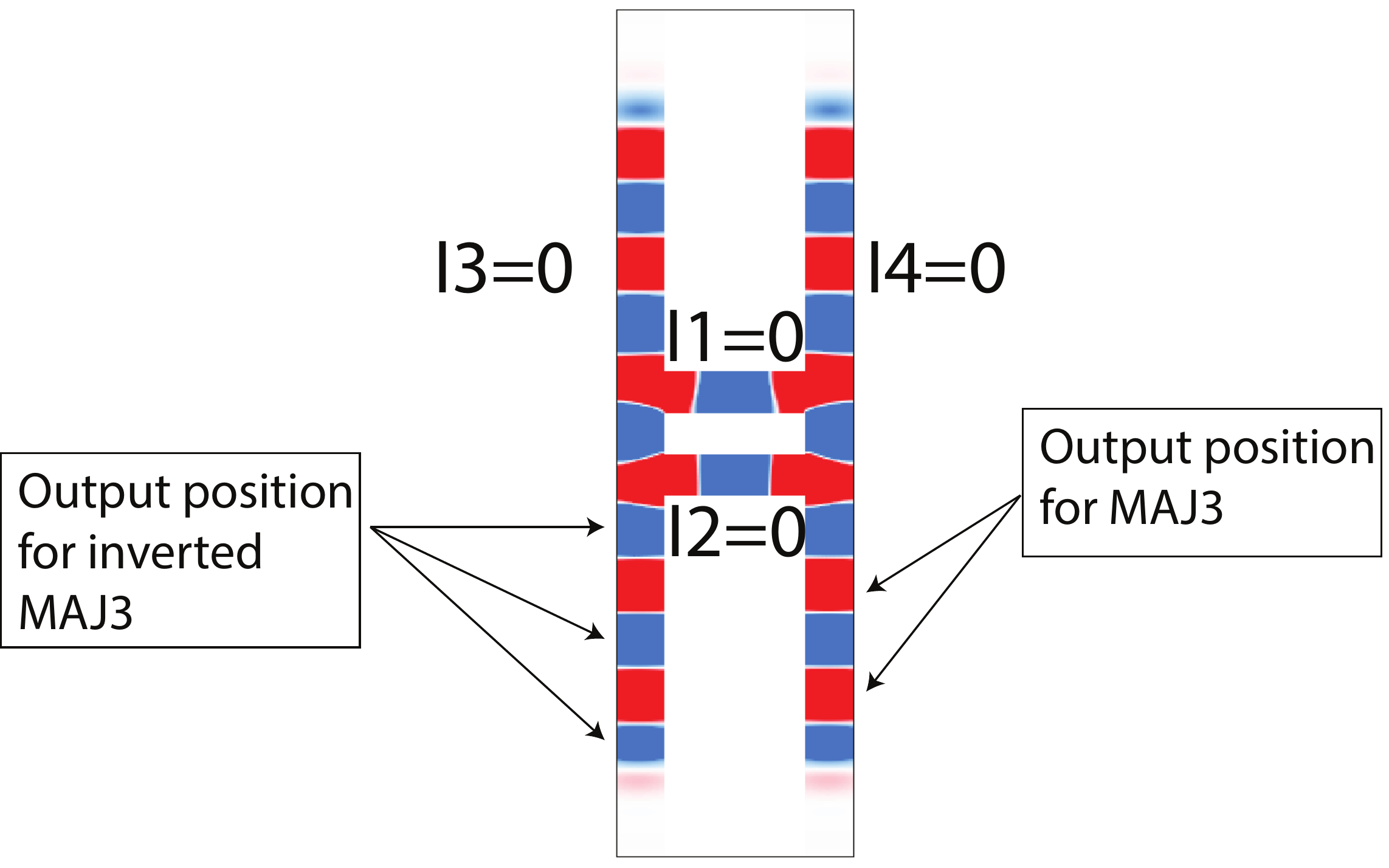}
  \caption{Inverted Outputs $O_1'$ and $O_2'$ and Non-inverted Outputs $O_1$ and $O_2$.}
  \label{fig:results1}
\end{figure} 

Figure \ref{fig:results1} presents the possibility of having the inverted and non-inverted outputs by adjusting the reading position. As one can observe in Figure \ref{fig:results1}, the inverted output ($O_1'$ and $O_2'$) of the Majority gates can be obtained by just shifting the reading position to a $\frac{n}{2} \lambda$ position.

By post-processing the OOMMF simulations, we estimated the MAJ3 gate delay, i.e., the maximum time it takes for the inputs to propagate to the output, as $1.5$ ns and $1$ ns for the $w = 75$ nm and $w = 50$ nm structures, respectively. To investigate the waveguide width reduction influence on SW group velocity Vg, we calculated the group velocities from micromagnetic simulation and obtained $Vg_{50nm} = 1.15$ $\mu$m/ns and $Vg_{75nm} = 1$ $\mu$m/ns for $w = 50$ nm and $w = 75$ nm structures, respectively. We also note that SWs are traveling shorter distances for the smaller structure, e.g., distance $I_3$ to $O_1$ is $380$ nm for $w = 50$ nm and $570$ nm for $w = 75$ nm. This imply that the $I_3$ to $O_1$ propagation takes $330$ ps for $w = 50$ nm and $570$ ps $w = 75$ nm. Therefore, gate performance increase is a consequence of both shorter travelling distance and increased group velocity. Thus, the gate delay can be further reduced by scaling down $w$, but also by making use of other waveguide materials.

\begin{figure}[t]
\centering
  \includegraphics[width=0.45\linewidth]{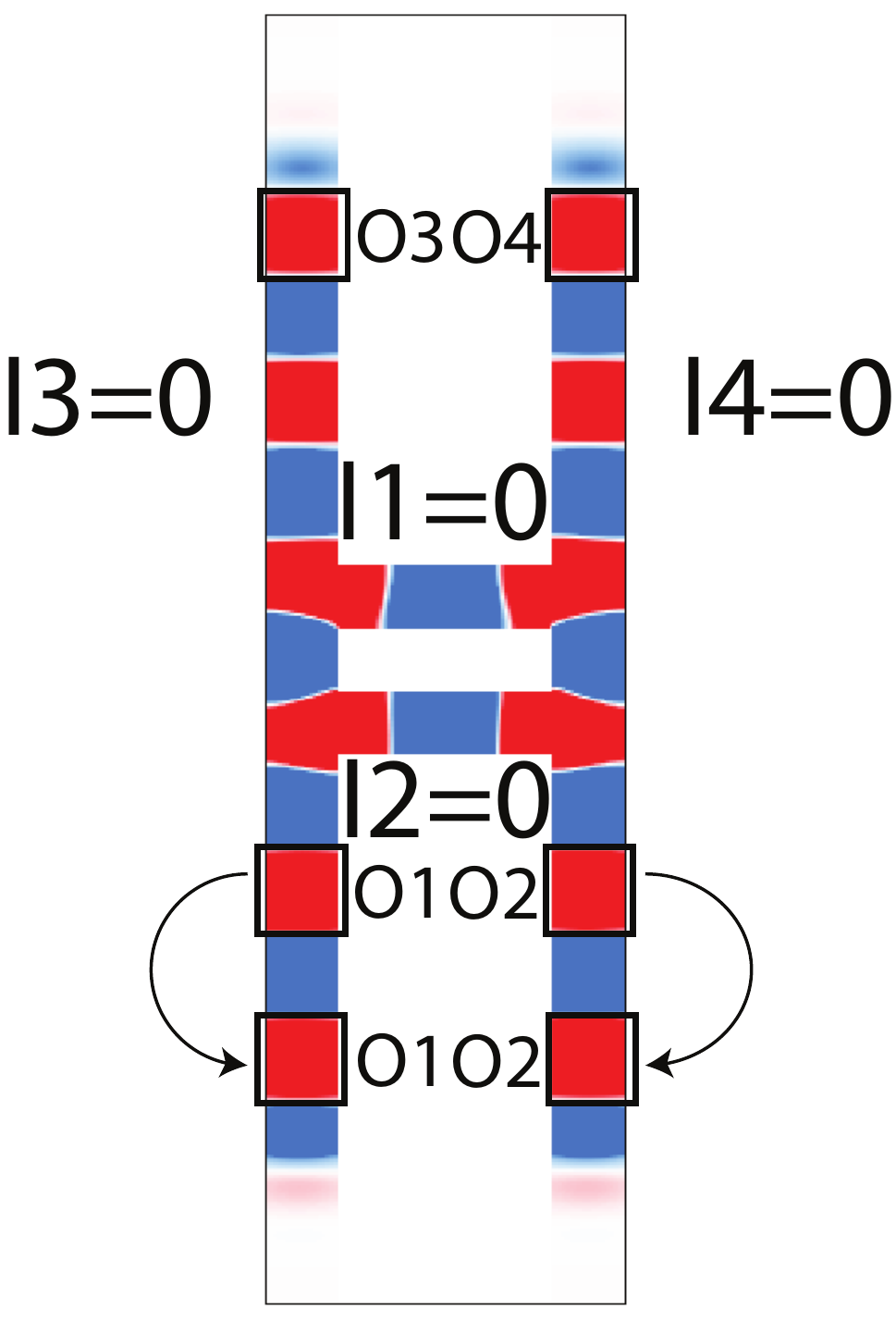}
  \caption{FO4 MAJ3 Gate.}
  \label{fig:FO4}
\end{figure}

We note that if only one MAJ3 output is required the structure can be simplified: i) physically, by removing one of its vertical waveguides (arms) or ii) logically, by not providing an input signal to $I_4$. Moreover, the gate fanout capabilities can be extended beyond $2$ by vertically lengthening its arms. For example, if the outputs in Figure \ref{fig:results} and \ref{fig:results1} are shifted downward to the end of the arms and two outputs are placed upward (at the upper-end of the arms), four outputs can be accommodated and if properly designed the gate can provide a fanout of $4$ as indicated in Figure \ref{fig:FO4}. However, the detailed design of such a structure constitutes future work and is out of the scope of the current paper.

To get inside on the quality of the achieved fanout, i.e., the similarity between the two SWs obtained at the gate outputs, we make use of Magnetization Spinning Angle (MSA) as metric. The input and output spinning angles are calculated as:
\begin{equation}
MSA = \arctan \left(\frac{\sqrt{(\overline{m_x})^2+(\overline{m_y})^2}}{M_s}\right),
\end{equation} 
where $\overline{m_x}$ and $\overline{m_y}$ are the x and y component of the magnetization, respectively.

Table \ref{table:2} presents the contribution percentage of each input to the outputs $O_1$ and $O_2$ when each of them is separately activated, for the $50$~nm  waveguide width design. The outputs MSAs in the Table are normalized value with respect to the activated input MSA. Thus, when only $I_1$ is activated $O_1$ and $O_2$ MSAs are normalized by $I_1$ MSA. The same holds true for the other 3 situations presented in the Table. As it can be noticed, $I_3$, $I_2$, $I_1$, and $I_4$ contributions to $O_1$ and $O_2$ are quite different. Due to symmetry, $I_1$ equally contributes to both gate outputs $O_1$ and $O_2$ and the same hold true for $I_2$ also. However, due to its proximity $I_2$ has a larger contribution to the outputs than $I_1$ and as such their strengths have to be properly balanced. Input $I_3$ SW is the strongest contributor to $O_1$ as it has a direct path to $O_1$, while spin waves from  $I_2$ and $I_1$ are facing edges, and reflect back and forth. 
Moreover, $I_3$ mostly affects $O_1$ and to a lower extend $O_2$, while $I_4$ effect is stronger on $O_2$ and weaker on $O_1$. Thus, as the inputs on the vertical and horizontal waveguides differently contribute to the outputs, $I_3$ and $I_4$ SWs must be excited at lower energy than $I_1$ and $I_2$ SWs to enable the correct gate behavior. 

Table \ref{table:3} presents the normalised (with respect to $I_1$) MSA of the outputs when all inputs are activated together for the same $w=50$ nm design.  
As it can be noticed from Table \ref{table:3}, the normalized $O_1$ and $O_2$ MSA is the same in all cases, which means that the proposed MAJ3 gate can successfully achieve a fanout of $2$.  One can also observe in the Table that different input combinations are producing different normalized MSA values. When all gate inputs have the same value ($I_1=I_2=I_3$), the output MSA is reaching the highest value because of the constructive interference. When inputs have different values, the destructive interference diminishes the spin wave energy, which results in lower MSA values. Moreover, when the horizontal inputs ($I_1$ and $I_2$) are different, the position of the asserted input affects the MSA output. For example, when ($I_3=1$, $I_2=0$, and $I_1=1$) or ($I_3=0$, $I_2=1$ and $I_1=0$) the the normalized output MSA is higher than when ($I_3=1$, $I_2=1$, and $I_1=0$) or ($I_3=0$, $I_2=0$, and $I_1=1$) because $I_2$ is located further than $I_1$ and $I_3$ from the interference location. As a result, when $I_1$ and $I_3$ have the same state, they interfere constructively and then destructively with $I_2$, which results in a larger magnetization angle.

\begin{table}[t]
\caption{Input Contribution Percentage on the Outputs - Separately Activated Inputs.}
\label{table:2}
  \begin{tabular}{|>{\centering}m{3em}|>{\centering}m{5em}|>{\centering}m{5em}|>{\centering}m{5em}|>{\centering}m{5em}|}
    \hline
    Inputs & $O_1/I$ \%  & $O_2/I$  \% \tabularnewline 
    \hline
    $I_1$ & $54$ \%  & $54$ \%  \tabularnewline
    \hline
    $I_2$ & $57$ \%  & $57$ \%  \tabularnewline
    \hline
    $I_3$ & $96$ \% & $35$ \%  \tabularnewline
    \hline
    $I_4$ & $35$ \% & $96$ \% \tabularnewline
    \hline
  \end{tabular}
\end{table}

\begin{table}[t]
\caption{Normalized Outputs ($O_1$ and $O_2$) by $I_1$ - Simultaneously Activated Inputs.}
\label{table:3}
  \begin{tabular}{|c|c|c|c|c|}
    \hline
   \multicolumn{3}{|c|}{Inputs} & $O_1/I_1$  \%  & $O_2/I_1$  \% \\ \hline
    $I_1$ & $I_2$ & $I_3$ AND $I_4$ &  &   \\
    \hline
    $0$ & $0$ & $0$ & $1$  & $1$ \\
    \hline
    $0$ & $0$ & $1$ & $0.28$ &  $0.28$ \\
    \hline
    $0$ & $1$ & $0$ & $0.37$ &  $0.37$ \\
    \hline
    $0$ & $1$ & $1$ & $0.45$ &  $0.45$ \\
    \hline
    $1$ & $0$ & $0$ & $0.45$ &  $0.45$ \\
    \hline
    $1$ & $0$ & $1$ & $0.37$ &  $0.37$ \\
    \hline
    $1$ & $1$ & $0$ & $0.28$ &  $0.28$ \\
    \hline
    $1$ & $1$ & $1$ & $1$ &  $1$ \\
    \hline
  \end{tabular}
\end{table}

An accurate evaluation of the proposed structure is not possible at this stage of development, especially for the energy and delay. That is mostly due to the missing excitation and detection cells figure of merit data. Thus, as the transducers are the dominant source for energy and delay, we chose to use the area as a metric to position our proposal versus existing state of the art. 

In order to make a fair comparison with \cite{Excitation_table_ref16}, we scaled down the MAJ3 design for $w = \lambda = 48$ ~nm and validate it by means of OOMMF simulations. In addition, the outputs are captured directly at the last interference point. The proposed scaled FO2 MAJ3 gate requires a real estate of $0.0576$ $\mu m^2$.  As the gate in \cite{Excitation_table_ref16} cannot provide fanout, we have to consider two such gates working in parallel on the same input set to evaluate both gates in similar utilization conditions, which results in a required area of $0.0691$ $\mu m^2$, i.e., our proposal provides a $16$ \% area reduction at the gate level. We note, however, that at the circuit level the area savings are significantly more substantial, as in order to deal with a fanout of $2$ gate output $O$ the approach in \cite{Excitation_table_ref16} requires the replications of all the gates on $O$'s cone of influence starting from the circuit primary inputs, and that for efficient logic synthesis of practical circuits gates with $ > 1$ fanout are frequently necessary. 

In order to compare with CMOS, we evaluated a $3$-input Majority gate implemented in $15$ nm technology with two NAND gates and one OR-AND-Invert (OAI) gate, at $V_{dd} = 0.8$ V, $25^{\circ}C$, and an output load  capacitance of $20$ fF. Our evaluation indicate that the $15$ nm CMOS MAJ3 area is $0.688$ $\mu m^2$, thus a $12$x larger area than the proposed SW MAJ3 gate.

In summary, we presented a novel fanout of $2$  area efficient $3$-input spin wave Majority gate (MAJ3).  We validated two instances of our proposal by means of OOMMF simulations and evaluated the fanout quality by making  use of  the Magnetization Spinning Angle (MSA) as metric. We calculated the normalized MSA values for the gate outputs and obtained negligible mismatch between them under all possible input combinations, i.e., a high quality fanout. We compared our proposal with MAJ3 SW, under the same material assumptions and utilization conditions, and $15$nm CMOS state of the art counterparts in terms of area and demonstrated a $16$ \% and $12$x less area, respectively. As a closing remark, we note that achieving $>1$ fanout is an enabling factor for the realization of SW circuits, as it eliminates the otherwise required circuit replication associated with fanout nodes intrinsic to SW circuits produced by means of logic synthesis. Thus, the implications of our proposal at the circuit level are a lot more substantial than at the gate level, both in terms of area and energy consumption.

\begin{acknowledgments}
\noindent
"This work has received funding from the European Union's Horizon 2020 research and innovation program within the FET-OPEN project CHIRON under the grant agreement No. 801055."
\end{acknowledgments} 

\vspace{0.5cm}
\noindent \textbf{Data Availability Statement}\\ \\
The data that supports the findings of this study are available within the article.

\nocite{*}
\bibliography{references}

\end{document}